\documentclass[authoryear,11pt]{article}
\title{
Protein Synthesis- Degradation,\\
 A Stochastic Approach-part I
}

\begin{document}
\author{Diana David-Rus $^{*}$\\
\\
$^{*}$National Institute for Physics and Nuclear Engineering (IFIN-HH)\\
 \small{R-77125,Bucharest-Magurele, Romania}}

\date{}
\maketitle

\begin{abstract}
In this work, we study a protein synthesis degradation process by defining a general  mathematical model. Using generating function technique we present a method that allows exact calculation of joint probability distribution 
 of protein copies in a cell for a two dimensional birth-death process with interaction.  We discuss the model in  steady state  for a particular choice of  transition  rules  and find exact solutions. 

\end{abstract}


\section{Introduction}

Proteins are essential macromolecules that serve both as structural components of the cell and as its enzymatic machinery. When stochasticity in this processes is ignored  the deterministic Michaelis-Menten model which can be understood also as a mean field approach is a good approximation. In most cases stochasticity  plays a significant role in the process that can not be ignored.  For example gene expression in both prokaryotes and eukaryotes is inherently
stochastic [1, 2]. 

Protein synthesis is a tightly gene expression regulated cellular process that affects growth, reproduction, aging and survival in response to both intrinsic and extrinsic cues, such as nutrient availability and energy levels. The turnover of these proteins (synthesis and degradation) is a stochastic process that plays a critical role in all biological processes. 


The aim of this work is to explore minimal models of protein synthesis- degradation and gain some analytical insight. For clarity of exposition,
in next section we present the analytic tools used to find the steady state probability distributions
of protein copy numbers in the cell and introduce the model in the general form.  After that I present the model for a particular case.


 Given that biochemical processes frequently involve small numbers of molecules  and such reactions are subject to significant stochastic fluctuations we consider that the minimal model considered here is relevant to such context  and therefore important to be studied at analytical level.
 
\section{Methods and general framework of the model}

We envision a protein synthesis degradation process as a continuous in time birth death Markov process with a discrete (very large) state space. When two different species or types of proteins with a large number of possible states are involved  the stochastic model that describes the process can be considered a two dimensional birth death Markov process. The interaction between the two protein types, I and II is "felt" in the creation rate of the protein type II as described bellow.
   The master equation gives the flow for $\pi(j, k; t)=\pi_{jk},$ the probability
of there being j copies of the 1st species, k copies of the second, at time
t.
 For a simple case where each reaction either creates or annhilates one and
only one component, and for the simple case where birth/decay rates
are constant, we have the following master equation describing the time evolution of the probability distribution

\begin{equation}
\frac{d\pi_{jk}}{dt}=\sum_{j=1}^{\infty}\sum_{k=1}^{\infty}(\pi_{j-1 k}\beta+ \pi_{j k-1}b_{jk}-\pi_{j k}j\delta-\pi_{j k}kd) 
\end{equation}  

where:
$\pi_{jk}$ is the joint probability distribution of type I and II to have j respective k quantities;
$\delta$ and d are the rates of annihilation of protein type 1 respectively 2 which are some constants proportional to the existing number of protein copies. The rates of creation for protein type 1 with j copies is $\beta$ and for protein type 2 with k copies is $b_{jk},$. 

The two protein types interact in the following way: when protein type 1 gets to a certain threshold $\theta$, protein type 2 is changing the creation/birth  rate according with bellow step function: 
     \[ b_{jk} = 
\left\{\begin{array}{cl}
          b_0, & \mbox{when $ j < \theta$} \\
          b_1  & \mbox{when $ j \geq \theta$} \\
        \end{array}\right.\]

We will study analytically the stochastic model formulated above in the steady state case for a particular choice of states and rules of state transitions.
 
Solving  master equation (1) analytically for the long time behavior of $\pi_{jk}$ is generally an impossible task when the state space is very large. One, therefore, has to resort on various techniques. One such technique often used successfully in stochastic processes literature  is the ``generating function technique" [18-20]. We remind the reader of some well-known aspects of this technique in order to make the present discussion self-contained. 
 
Assume j is a discrete random variable and assume, for convenience, the state space is $\{0,1,2...\}$.
Let $\pi_j$ be the probability mass function of j where $\sum_{j=0}^{\infty}\pi_j=1$; the mean of j satisfy:
$\mu_j=E(j)=\sum_{j=0}^{\infty} j\pi_j.$
The probability generating function (p.g.f.) of the discrete random variable j is  defined by \[f_j(x)=E(x^j)=\sum_{j=0}^{\infty}\pi_j x^j\] for some $x\epsilon {R},$ because $\sum_{j=0}^{\infty}\pi_j=1$, the above sum converges absolutely for $\vert x\vert \leq 1.$ As the name implies, the p.g.f. generates the probabilities associated with the distribution, where 
$f_{j}(0)=\pi_0,$ $f'_{j}(0)=\pi_1,$ $f"_{j}(0)=2!\pi_2,$ and in general $f^{n}_{j}(0)=n!p_n.$ The p.g.f. gives entire information associated with the distribution.

\section{Minimal model considered: Two state model}
Given that real biological systems frequently involve small numbers of molecules we developed a minimal model  where the second protein type can be in 2 possible states: present or absent. An example of such situation in real life would be a genetic switch on/off.

The two protein types  undergo a birth/death process with interaction.  The creation and anihilation rates are the same as in general model with the following specifications:
 type I protein can have any number of copies/states, while type II protein can only have 
2 possible copies/states: 0 or 1, meaning we have no  protein  or just  one 
protein. The interaction between the 2 protein types is as following: the creation rates of the second protein type, $b_{jk},$ will directly depend by the number of copies of first protein type as bellow: 
 

     \[ b_{jk} = 
\left\{\begin{array}{cl}
          b_0, & \mbox{when $ j \leq \theta$} \\
          b_1  & \mbox{when $ j \geq \theta$} \\
          0& \mbox{when $ k > 1 \forall j$} 
        \end{array}\right.\]

 For simplicity we will chose $\theta =1$ but the result can be easily generalized, see fig. 2 (with arrows and transition rates) 


For this model, in  steady state the master equation (1) is replaced with bellow equations, where we keep the same notation as in equation (1).
 
\begin{eqnarray}
&
  j = 0, k = 0:
&\pi_{00}(\beta +b_0) = \pi_{10}\delta + \pi_{01}d,   \label{2.1}\\
& 
j \geq 1, k = 0: 
& \pi_{j0}(j\delta + \beta + b_1) = \pi_{j+1,0}(j+1)\delta +\pi_{j1}d +\pi_{j-1,0}\beta,   \label{2.2}\\
& 
j = 0, k = 1:
&\pi_{01}(\beta + d) = \pi_{11}\delta + \pi_{00}b_0, \label{2.3}\\
&
j \geq 1, k = 1:
&\pi_{j1}(j\delta + \beta + d) = \pi_{j+1,1}(j+1)\delta +\pi_{j0}b_1 + \pi_{j-1,1}\beta, \label{2.4}
\end{eqnarray}
Using generating function technique we simplify  our problem by transforming the above equations into ODE's satisfied by the generating function.
Let $f_k(x) = \sum_{j=0}^{\infty} \pi_{jk}x^j$ be the probability generating function (p.g.f); the differential of p.g.f.is:$f_k'(x)= \sum_{j=0}^{\infty} j\pi_{jk}x^{j-1};$
 where $ \sum_{j=0}^{\infty} \pi_{jk}x^j$ converges absolutely for $\mid x\mid \leq 1.$ According with the above notation we  have:
$f_k(0) = \pi_{0k}$, $ f_0(0) = \pi_{00} $,  $f_1(0)=\pi_{01},$
and $ f_0(x) = \sum_{j=0}^{\infty} \pi_{j0}x^j,$ $ f_1(x) = \sum_{j=0}^{\infty} \pi_{j1}x^j.$  

Given that p.g.f. generates the probabilities associated with the distribution as previously explained, the problem is to find an analytical expression for $f_0(0)$ since once we would know such expression, we have all the necessary information to know the joint probability distribution for the 2 protein types. 

Note that for the case when we have just one protein type undergoing a birth death process with a constant decay/birth rate is a well known fact that in steady state, its  stationary probability distribution  is a Poisson distribution [19,20]. Since protein 2 doesn't influence protein 1, the marginal distribution of protein 1 is still given by the Poisson distribution:
$p_j = \pi_{j0} + \pi_{j1}= \frac{1}{j!}\left(\frac{\beta}{\delta}\right)^j e^{-\beta/\delta}.$ It follows that marginal, in the generating function notation is correct to right: 
 \begin{equation}
f_0(x)+ f_1(x)=\sum_{j=0}^{\infty} x^j(\pi_{j0} + \pi_{j1}) = \sum_{j=0}^{\infty} x^je^{-\beta/\delta}\frac{1}{j!}\left(\frac{\beta}{\delta}\right)^j=\\
e^{-\beta/\delta}\sum_{j=0}^{\infty}\frac{1}{j!}\left(x\frac{\beta}{\delta}\right)^j=
e^{-\beta/\delta}e^{x\beta/\delta}=\]  
\[=e^{(x-1)\beta/\delta}
\end{equation}  
and
\begin{equation}
f_0^{'}(x)+f_1^{'}(x)=\frac{\beta}{\delta}e^{(x-1)\beta/\delta}
\end{equation}

Using generating function technique on equations 2,3  an ODE equation (eq.8) satisfied by a generating function is derived (see Appendix A for a detailed derivation)  

\begin{equation}
 x\delta\frac{d}{dx} f_0(x) + (\beta+b_1)f_0(x) + (b_0 - b_1)f_0(0)=\delta\frac{d}{dx} f_0(x) + f_1(x)d + f_0(x)\beta x  
\end{equation}
same procedure applied on eq 4,5 and obtain the following equation: 
\begin{equation}
 x\delta\frac{d}{dx} f_1(x) + (\beta+d)f_1(x) =  (b_0 - b_1)f_0(0)+\delta\frac{d}{dx} f_1(x) + f_0(x)b_1 + xf_1(x)\beta 
\end{equation}  
\textbf{Steps toward obtaining  $f_0(x):$}
Using condition (6) in eq.(8) I  obtain a new equation (eq.10)in the $f_0(x)$ as unknown which once solved gives me 
the expression for $f_0(x)$ generating function. 
\begin{equation}
(x-1)\delta\frac{d}{dx} f_0(x) +(-\beta(x-1)+b_1+d)f_0(x)=de^{(x-1)\beta/\delta}+(b_1-b_0)f_0(0)
\end{equation}
This is a first order  ODE;
 Using integrand factor method one gets after some calculations (see Appendix B for a detailed derivation in solving eq.10)

\begin{equation}
f_0(x)(1-x)^{(b_1+d)/\delta} e^{-\beta/\delta x}-f_0(0)=\]\[-de^{-\beta/\delta}\frac{1}{b_1+d}(1-(1-x)^{(b_1+d)/\delta})-\frac{(b_1-b_0)f_0(0)}{\delta}\int_{0}^{x}e^{-\frac{\beta}{\delta} y}(1-y)^{\left(\frac{b_1+d}{\delta} -1\right)}dy
\end{equation}
Setting $x=1$ in equation 11 one obtains $f_0(0)$ and than going back at eq.11 one gets an expression  for $f_0(x)$ 
\section{RESULTS}
Setting $x=1$ in eq. (9) one obtains $f_0(0)$:
\begin{equation}
f_0(0)=\frac{de^{-\beta/\delta}}{b_1+d}\left[\frac{1}{1-\frac{b_1-b_0}{\delta}
\int_{0}^{1}e^{-(\beta/\delta) y}(1-y)^{\frac{b_1+d}{\delta} -1}dy}\right]
\end{equation} 
Given that the derivatives of generating fct at zero gives the probabilities associated with the distribution, we have:

 \[\pi_{00} = f_0(0)\] 
\[\pi_{10}=f'_0(0)\]
\[\pi_{20}=f''_0(0)\]
\[..................\]
\[\pi_{n0}=f^n_0(0)\]

From  (6) we have $f_1(x)=e^{(x-1)\beta/\delta}-f_0(x)$ and then for $x=0$ one obtains: 
 $f_1(0)=e^{-\beta/\delta}-f_0(0)$

therefore,in same way I can easily get: 

 \[\pi_{01} = f_1(0)\] 
\[\pi_{11}=f'_1(0)\]
\[\pi_{21}=f''_1(0)\]
\[..................\]
\[\pi_{n1}=f^n_1(0)\]
\section{Discussions and conclusions}
Using the  result above one can determine the joint probability distribution   of having a given number of proteins type I and II in the system in the stationary state, given that the  protein type II can have just 2 possible states.
 
Using generating function technique I've shown how one can get a close analytical expression for joint probability distribution for the particular choice of states.

\section*{Acknoledgements}
This work would have not been possible without the guidance of Prof. Larry Shepp and the valuable insights of Prof. Joel L. Lebowitz. Note that this result was previously comunicated to Dr. Chris Wiggins who challenged us with the problem of  identifying an analytical expression. \\

 This work was partly supported by  NSF Grant DMR-0802120 and AFOSR Grant AF-FA-9550-04-4-22910. Also by POSDRU/89/1.5/S/60746.21 and CNCSIS-UEFISCDI, project numberPN-II-PT-PCCA-2011-3.1-1350 

\section {References}

[1] Maheshri, O'Shea, E.K., Annu. Rev. Biophys. Biomol. Struct. 36,413 (2007).

[2] Elowitz MB, Levine AJ, Siggia ED, Swain PS, Science 297,1183 (2002)

[3] Blake WJ, Kaern M, Cantor CR, Collins JJ, Nature 422, 633 (2003)

[4] Raser JM, O'Shea EK, Science 304,1811(2004)

[ 5] Kaern, M., Elston, T.C., Blake, W.J. Collins, J.J. Nat. Rev. Genet. 6, 451 (2005).

[6] Kaufmann, B.B.  van Oudenaarden, A, Curr. Opin. Genet. Dev. 17, 107 (2007).

[7] J. S. van Zon and P. R. ten Wolde, Physical Review Letters 94, 128, 103  (2005).

[8] B. Munsky and M. Khammash, Journal of Computational Physics 226, 818 (2007).

[9] D. Gillespie, M. Roh, and L. Petzold, J. Chem. Phys. 130, 174103 (2009).

[10] Brian Drawert, Michael J. Lawson, Linda Petzold, and Mustafa Khammash, J. Chem. Phys. 132, 074101 (2010)

[11]  Bailey N.T.J, The Elements of Stochastic Processes with Applications to the Natural Sciences, J. Wiley and Sons, New York, (1990) 

 [12]  Karlin,S. and H.Taylor, First and Second Course in Stochastic Processes, (1975, 1981)

 [13] Linda S. Allen,Stochastic processes with applications to Biology, (2003)

[14] Aleksandra M. Walczak, Andrew Mugler, Chris H. WIggins, arXiv:1005.2648 [q-bio.MN] or arXiv:1005.2648v1 [q-bio.MN]

\pagebreak

\section*{Appendix}
\appendix 

\section{Apply generating function technique  in two state model for obtaining an ODE satisfied by g.f.: eq. (8)}

Using generating function technique on equations 2,3 (in equation 2+3 multiply by $x^j$  and sum over j from 0 to $\infty$ each term) an ODE equation (eq.8) satisfied by a generating function is derived.
  \\ 
In equation 2+3 multiply by $x^j$  and sum over j from 0 to $\infty$ each term.

  \[ \delta \sum_{j=1}^{\infty} j\pi_{j0}x^j + \sum_{j=1}^{\infty} (\beta +b_1) \pi_{j0}x^j + \beta\pi_{00} + b_0\pi_{00} =\]  \[\delta \sum_{j=1}^{\infty}  
 (j+1)\pi_{(j+1),0}x^j + d\sum_{j=1}^{\infty} \pi_{j1}x^j +\beta\sum_{j=1}^{\infty}\pi_{(j-1),0}x^j+ \delta\sum_{j=0}^{\infty} \pi_{10}x^j 
+ d \sum_{j=0}^{\infty} \pi_{01}x^j\].

where each term in the equation can be written:

I  \[\delta \sum_{j=1}^{\infty} j\pi_{j0}x^j = x\delta\sum_{j=0}^{\infty} j\pi_{j0}x^{j-1} = x\frac{d}{dx}f_0(x) \delta\].  

II \[\sum_{j=1}^{\infty} (\beta +b_1) \pi_{j0}x^j = \sum_{j=0}^{\infty} (\beta +b_1) \pi_{j0}x^j - (\beta +b_1)\pi_{00}x^0 = (\beta +b_1) f_0(x) - (\beta +b_1) f_0(0)\]. 

III \[\beta\pi_{00} = \beta f_0(0),\] 

IV   \[ b_0 \pi_{00} = b_0 f_0(0)\]

V  \[\delta \sum_{j=1}^{\infty} (j+1)\pi_{(j+1),0}x^j = \delta \sum_{j=0}^{\infty} (j+1)\pi_{(j+1),0}x^j -\delta\pi_{10} \]

VI  \[d\sum_{j=1}^{\infty} \pi_{j1}x^j = d \sum_{j=1}^{\infty} \pi_{j1}x^j-d\pi_{01}= f_1(x) - d\pi_{01}\]

VII  \[\beta\sum_{j=1}^{\infty}\pi_{(j-1),0}x^j = \beta x f_o(x)\]

VIII    $\delta\pi_{10}$ ,   IX   $d\pi_{01}$

introducing all terms from I- IX  back in equation we obtain 

  equation (8):

\[ x\delta\frac{d}{dx} f_0(x) + (\beta+b_1)f_0(x) + (b_0 - b_1)f_0(0)=\delta\frac{d}{dx} f_0(x) + f_1(x)d + f_0(x)\beta.\]  

Similarly, applying generating function technique on eq 4,5 yields the following equation: 

(9)
\[
 x\delta\frac{d}{dx} f_1(x) + (\beta+d)f_1(x) =  (b_0 - b_1)f_0(0)+\delta\frac{d}{dx} f_1(x) + f_0(x)b_1 + xf_1(x)\beta 
\] 

Further, using expression (6) in eq.(8) I  obtain  equation (eq.10) for $f_0(x)$ 
(10)
\[
(x-1)\delta\frac{d}{dx} f_0(x) +(-\beta(x-1)+b_1+d)f_0(x)=de^{(x-1)\beta/\delta}+(b_1-b_0)f_0(0)
\nonumber
\]


\section{ Derivations of eq.8 in two state model-Integrating factor method}
(8)  \[(x-1)\delta\frac{d}{dx} f_0(x) +(-\beta(x-1)+b_1+d)f_0(x)=de^{(x-1)\beta/\delta}+(b_1-b_0)f_0(0)\]

 multiply eq. (8) with the integrand factor:
\[\left[(x-1)\delta f_0^{'}(x) +(-\beta(x-1)+b_1+d)f_0(x)\right]e^{\int_{0}^{x}\frac{b_1+d-\beta(y-1)}{(y-1)\delta}dy}=\]
\[(x-1)\delta\frac{d}{dx}\left( f_0(x)e^{\int_{0}^{x}\frac{b_1+d-\beta(y-1)}{(y-1)\delta}dy}\right)=\left(de^{(x-1)\beta/\delta}+(b_1-b_0)f_0(0)\right)e^{\int_{0}^{x}\frac{b_1+d-\beta(y-1)}{(y-1)\delta}dy}=\]

where:
\[\int_{0}^{x}\frac{b_1+d-\beta(y-1)}{(y-1)\delta}dy = -\frac{\beta}{\delta} x +log(1-x)^{(b_1+d)/\delta}\] 

therefore:
\[e^{\int_{0}^{x}\frac{b_1+d-\beta(y-1)}{(y-1)\delta}dy}=(1-x)^{(b_1+d)/\delta} e^{-(\beta/\delta) x}\]

from here we get by dividing with $(x-1)\delta$:
\[\left(\frac{d}{dx}f_0(x)e^{\int_{0}^{x}\frac{b_1+d-\beta(y-1)}{(y-1)\delta}dy}\right)= 1/(x-1)\delta\left(de^{(x-1)\beta/\delta}+(b_1-b_0)f_0(0)\right)e^{-(\beta/\delta) x} e^{\left((b_1+d)/\delta\right) log(1-x)}\]

Integrating this eq. we obtain:

(10) \[f_0(x)e^{\int_{0}^{x}\frac{b_1+d-\beta(y-1)}{(y-1)\delta}dy} - f_0(0) = 
-\int_{0}^{x}\left(de^{(y-1)\beta/\delta}+(b_1-b_0)f_0(0)\right)e^{-(\beta/\delta) y}(1/\delta) 1/(1-y)^{(1-(b_1+d)/\delta)}\] 

or equation. (11.a):

\[f_0(x)(1-x)^{(b_1+d)/\delta} e^{-\beta/\delta x}-f_0(0)=I_1 +I_2;\]  

where \[I_1=-d/\delta\int_{0}^{x}e^{\beta/\delta(y-1)}e^{-(\beta/\delta)y}\frac{1}{(y-1)^{((-b+d)/\delta)+1}}dy=\]
\[=-de^{-\beta/\delta}\frac{1}{b_1+d}(1-(1-x)^{(b_1+d)/\delta})\]

and 
\[I_2=-\frac{(b_1-b_0)f_0(0)}{\delta}\int_{0}^{x}e^{-\frac{\beta}{\delta} y}(1-y)^{\left(\frac{b_1+d}{\delta} -1\right)}dy\]

by introducing the expression $I_1 $ and $I_2$ in equation 11 we obtain the expression for $f_0(x)$:

[11]

\[f_0(x)(1-x)^{(b_1+d)/\delta} e^{-\beta/\delta x}-f_0(0)=\]\[-de^{-\beta/\delta}\frac{1}{b_1+d}(1-(1-x)^{(b_1+d)/\delta})-\frac{(b_1-b_0)f_0(0)}{\delta}\int_{0}^{x}e^{-\frac{\beta}{\delta} y}(1-y)^{\left(\frac{b_1+d}{\delta} -1\right)}dy\]

\end{document}